\documentclass{aa}
\usepackage{graphicx}
\begin{document}
\thesaurus{02.08.1;02.13.2; 11.11.1; 11.19.2}
\title{Global three-dimensional simulations of magnetic field
evolution in a galactic disk}
\subtitle{II. Gas rich galaxies}

   \author{S.~von Linden \inst{1} \and K.~Otmianowska-Mazur \inst{2}
        \and   H.~Lesch \inst{3} \and  G.~Skupniewicz \inst{2}}
   \offprints{S. von Linden}
        \mail{svlinden@lsw.uni-heidelberg.de}

\institute{Landessternwarte Heidelberg, K\"onigstuhl, D-69117
Heidelberg, Germany
\and Astronomical Observatory, Jagiellonian University, ul.
                   Orla 171, Krak\'ow, Poland
\and Universit\"ats-Sternwarte, Scheinerstr. 1, D-81679 M\"unchen, Germany}

\date{Received; accepted}

\titlerunning{Global 3-D simulation of magnetic
                  field evolution. II}
\authorrunning{S. von Linden et al.}
   \maketitle
\begin{abstract}

A fully three-dimensional computation of induction processes in a
disk is performed to investigate 
the evolution of a large-scale magnetic field in gas-rich barred and
spiral galaxies in the presence of field diffusion.
As input parameters we use time
dependent velocity fields obtained from self-consistent 3D N-body
calculations of galactic dynamics.
Our present work primarily analyzes the
influence of the gas flows in non-axisymmetric gravitational disturbances
(like spiral arms and bars) on the behavior of the magnetic field.
The magnetic field is found to evolve towards structures resembling
dynamical spiral arms and bars, however its distribution is often much more
complicated than the velocity field itself. We show that a random
configuration of seed magnetic fields quickly dissipates,
and is therefore unable to explain the magnetic field
strength observed in nearby galaxies.
The detailed comparison between the simulated
velocity and magnetic fields allows us to establish that the reversals
of magnetic
vectors appear in the vicinity of the corotation radius of non-axisymmetric
disk structures, where the velocity field shows rather abrupt changes.
We argue that this explains the observed field reversals in our Galaxy.

\end{abstract}

\section{Introduction}
The origin and evolution of large-scale
magnetic fields in galaxies are still not properly understood
(e.g. Beck et al. 1996; Lesch \& Chiba 1997). There are several
issues which are not yet clear. For example, there is still no
consensus on the primordial origin of large-scale magnetic fields
as proposed by Kulsrud \& Anderson (1992). The
physical processes initiating and driving turbulent galactic dynamos
may not be able to excite large-scale fields because of the magnetic
back reaction on the small scale turbulence (e.g. Cattaneo and Vainshtein
1991). Furthermore, the standard dynamo approach for investigating
large-scale magnetic fields does not include the non-axisymmetric part of the
galactic velocity field which is able to explain several observed
features in barred galaxies (Chiba and Lesch 1994).
We concentrate on this second issue in this contribution,
which is an extension of our previous work (Otmianowska-Mazur et al. 1997,
Paper I).
In Paper I, we coupled, for the first
time, fully three-dimensional velocity fields resulting from dynamical
simulations, with the large-scale magneto-hydro-dynamical evolution.
Our main finding was an
extremely sensitive response by the magnetic field
to spatial variations of the gas velocity field.
We concentrated on barred
galaxies with normal gas content, where the 
magnetic vectors are found to show opposite 
directions along the sides of the bar, and in the course of 
disk rotation the field vectors are mixed at the ends of the bar.
We have also found that the magnetic field lines are well
aligned with the bar and the spiral arms.

In the present paper, we generalize our study to consider gas-rich spiral
galaxies. The influence of the gas mass and gas contents on the structure
and evolution of the magnetic field is discussed.

In the magnetic field simulations, we are interested
in galaxies which undergo significant dynamical evolution
in terms of bar and spiral arm excitation. It is well known that the
intensity of the corresponding
dynamical instabilities depends sensitively on the amount of gas in the
disk. The higher the gas content the faster and stronger
are the excited non-axisymmetric gravitational instabilities which drive
the spiral arms (Lynden-Bell \& Kalnajs 1972), as can be 
observed in actual galaxies (Rix \& Zaritsky 1995). It is also known that
the excitation of such non-axisymmetric structures in disks
leads to strong angular momentum
transport toward the galactic center (Roberts et al 1979, Toomre
1977, Fuchs \& von Linden 1997).
The angular momentum is either removed from or transferred
to the stellar population. This process develops
the non-axisymmetric gravitational potential (Schwarz 1981).
Since
magnetic fields are strongly coupled with gaseous velocity fields, especially
the velocity gradient,
the amount of gas in the disk will influence the results
for the magnetic field evolution as well. In order to study this influence
in the present work, we perform 
calculations with gas masses of 10\% and 20\% of the total galactic mass.
The simulation of a gas-rich galaxy
allows us to focus also on proto-galaxies which have not undergone
a complete transformation of primordial gas into stars.

\section{Models and numerical methods}

Massive galactic disks are able to form  strong elongated bars
(Sellwood 1981). This allows us to concentrate on
a highly anisotropic velocity distribution accompanied by
the development of a bar.
Since the growth time scale of a bar increases with the ratio of
the halo masses and the disk masses $M_h/M_d$ (Combes \& Sanders 1981)
we follow a long-lived strong bar by choosing
$M_{h+b}/M_d$ = 2 ($M_d$= stellar disk mass, $M_{h+b}$=
halo and bulge mass).
To fulfill the bar instability criterion
\begin{equation}
\Omega_p > (\Omega(r) - \kappa(r)/2),
\end{equation}
where $\Omega_p$ is the pattern velocity of the bar, $\Omega  (r)$ is
the angular velocity of the disk and $\kappa(r)$ the epicyclic frequency.
The disk has to be massive enough so
that the mass in the inner part of the disk reaches a critical
mass $ M > M_{crit}$ (Sellwood 1981).
We compare our simulations of a gas-rich barred galaxy (model III) with a
model presenting only spiral activity in a less massive disk (model IV).

\subsection{N-body numerical scheme}

Our simulations are performed with a 3D N-body code
involving a molecular cloud scheme as described by
Combes \& Gerin (1985) and Gerin et al. (1990).
The modeled galaxy is embedded in a spherical halo
consisting of visible and dark matter and in a bulge component in the
central part of the disk. The galactic disk consists
of gas clouds and stellar particles which
are interacting via gravity.

The stars interact only by softened gravitational forces,
where the softening length $\epsilon$ is much smaller than the
scale height of the
disk so that there is practically no artificial stabilization caused by
large $\epsilon$ parameters (Romeo 1994).
The gas clouds can interact inelastically, which
is simulated in an elaborate cloud--in--cell scheme
describing the coalescence and
fragmentation of the clouds. We adopt the same scheme and
local processes as described by
Casoli \& Combes (1982) and Combes \& Gerin (1985).
The clouds encompass masses between $5\cdot\,10^{2}$ to
$5\cdot\,10^{5}$ M$_\odot$ distributed in 10 mass-bins with
logarithmic intervals in order to take into account the power-law variation
of the mass spectrum.

Alternative methods  to simulate the gas component as a fluid
are smoothed particle hydrodynamics (e.g. Friedli \& Benz 1993)
or finite-difference schemes.
Describing the dynamical features in disks,
there is no qualitative difference between these codes.
In a future work, we will
use different schemes for the gas particles to check the influence
on the magnetic field.

We perform our simulation with $N_s\sim 38\,000$ stellar particles.
To investigate the influence of the gas
particle number on the galactic evolution
we perform  simulations with $N_g \le 38000$ and $\le 19000$.
We use the same grid size and FFT method for the calculation of the
gravitational potential as described in Paper I.

The time step $\delta t$ (=10$^6$\,yr) of the N-body simulation is chosen
to fulfill the
relation $v_{phi,max}\delta t \leq l$, where $v_{phi,max}$
is the maximum of the rotation curve and $l$ is the grid length
(Combes \& Sanders, 1981).
The time step for the cloud-cloud collision is about $10^7$\,yr.
Several different simulations have been performed, varying the input
parameters for the disk, the bulge, the number of gas particles
and the gas mass.

\subsection{Magnetic field evolution model}

We analyze the time dependent solutions
of the induction equation:
%%%
\begin{equation}
\partial \vec{B}/\partial t=\hbox{rot}(\vec{v}\times\vec{B})-\hbox{rot}
(\eta\,\hbox{rot}\vec{B}),
\label{eq2}
\end{equation}
%%%
\noindent
where $\eta$ is the magnetic diffusion coefficient,
$\vec{v}$ is the velocity of the gas
and $\vec{B}$ is
the magnetic field. As in Paper I, the computations are performed
with the ZEUS3D code (Stone \& Norman, 1992) using only that part which
deals with
magnetic field evolution. In the original form of the code the constrained
transport (CT) algorithm is
implemented for the flux-freezing approximation of the eq.(\ref{eq2})
($\eta=0$).
In our analysis, the effects of the magnetic diffusion are introduced in two
ways: directly into the code using the CT algorithm
and by "pulsed flow" method (Bayly \& Childress,
1989) described in Paper I. Both methods fulfill the condition $div{B}=0$.

To incorporate the particle-based velocity field into the grid-based code,
a spline function interpolation is applied (cf. Paper I).

The computations have been performed using Cray Y-MP in HLRZ J\"ulich and
Convex Exemplar SPP-1000 Series in ACK CYFRONET-KRAK\'OW.

\begin{table}
\begin{flushleft}
\begin{tabular}{lllllll}
\hline\noalign{\smallskip}
% model gal99, gal98 and model gal81
Model    &III      & IV \\
\hline\noalign{\smallskip}
Mass in $10^{10} {\rm M}_\odot$:     &        &\\
- dark-halo mass $M_h$               & 9.6 & 9.6\\
- disk mass    $M_d$                 & 7.2 & 5.8 \\
- gas mass  $M_g$                    & 1.6 (0.8)& 1.6 \\
- bulge mass   $M_b$                 & 4.8 & 6.2\\
- mass ratio $(M_d+M_g)/M_{\rm tot}$ & 0.38 (0.36)& 0.28\\
Scale length in kpc: &  &\\
- dark halo $h$      & 15 &15\\
- bulge      $b$     &1.1& 1.2\\
- disk  $d$          &7&6\\
\noalign{\smallskip}
\hline
\end{tabular}
\caption[]{\label{tab1}Input parameters for model III and IV}
\end{flushleft}
\end{table}

\section{Input parameters}

We study numerically two experiments
provided by our 3D particle-mesh scheme simulations: the barred galaxy
(case III) and the spiral galaxy (case IV).
For both the halo and the bulge components, Plummer spheres are adopted.
In all models, the mass and the scale
length of the dark matter halo are the same as in Paper I.
The model parameters for the N-body calculations are given in Table \ref{tab1}.

The calculations for the magnetic field evolution are performed in
3D rectangular coordinates, where the XY plane is the galactic plane and
the Z axis is the axis of galactic rotation. The rectangular size
in the X and Y directions is 30 kpc. As a grid interval, we choose the
distance of 300 pc resulting in 101 grid points along the X and Y axis.
The experiments performed in Paper I have revealed that simulations
with twice the distance between the grid points are significantly
affected by numerical dissipation due to truncation errors present in
the ZEUS3D code (see Paper I, $\, 3.2$ and $\, 4.3$). Therefore, we decide
to adopt a smaller interval which allows us to diminish the numerical
diffusion to some extent.
The scale height
of our model galaxy is 1~kpc with 50~pc as the grid interval (21 grid
points along the Z axis).
The model input parameters for the magnetic field evolution
are summarized in Table \ref{tab2}.
The adopted time step of $10^{5}$~yr fulfills the Courant condition for the
given values of the velocity field and the magnetic diffusion coefficient.

\begin{table}
\begin{flushleft}
\tabcolsep 0.1cm
  \begin{tabular}{lllllll} \hline\noalign{\smallskip}

Model & $dx, dy$ & $dz$ & $N_{x}\cdot N_{y}\cdot N_{z}$& $\eta$ & Input\\
      & [pc]   & [pc]&                              & [cm$^2$/s]&$B$-field&\\
\noalign{\smallskip}\hline\noalign{\smallskip}
IIIa    & 300 & 50& $101\cdot 101\cdot 21$& 0& toroidal&\\
IIIb   & 300 & 50& $101\cdot 101\cdot 21$&$3.2\cdot 10^{26}$&toroidal& \\
IIIc   & 300 & 50& $101\cdot 101\cdot 21$& $0$& random&\\
IIId   & 300 & 50& $101\cdot 101\cdot 21$& $3.2\cdot 10^{26}$&random& \\
\noalign{\smallskip}\hline\noalign{\smallskip}
IVa   & 300 & 50& $101\cdot\,101\cdot\,21$& $0$& toroidal&\\
IVb   & 300 & 50& $101\cdot\,101\cdot\,21$& $3.2\cdot 10^{26}$&toroidal& \\
\noalign{\smallskip}\hline
\end{tabular}
\end{flushleft}
\caption[]{\label{tab2} The magnetic field evolution model parameters}
\end{table}
\begin{figure}
%     \resizebox{\hsize}{!}{\includegraphics[angle=0]{gas.ps}}
\caption[]{Distribution of molecular clouds
for the model III projected onto the galactic plane
(at t= $3.5 \cdot\,10^8$\,yr.). Massive clouds are plotted with
thicker dots as less massive clouds.}
\label{bari}
\end{figure}
\begin{figure}
     \resizebox{\hsize}{!}{\includegraphics[angle=-90]{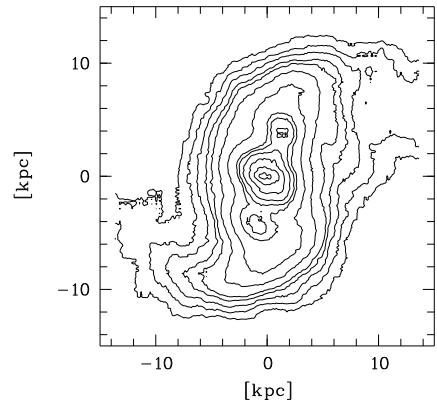}}
\caption[]{Isophotes of the stellar disk density
for the model III (at t= $3.5 \cdot\,10^8$~yr.).}
\label{iso}
\end{figure}

The computations were performed for two initial magnetic field
configurations: a purely axisymmetric toroidal field, going in the
anti-clockwise direction, and a random field.
The strength of the axisymmetric field
is 1 $\mu$G and decreases both toward the galaxy center and
toward the edges
(see Paper I for a more detailed description).
The random field is calculated numerically
from a random distribution of the magnetic vector potential $\vec{A}$.
Such a procedure
provides us with the magnetic field structure free of magnetic mono-poles.
The mean strength of the random field is about 1\,$\mu$G.
As in Paper I, we adopt "outflow" boundary conditions as the
most stable for our calculations. The assumed turbulent diffusion
coefficient is
$3.2\cdot\,10^{26}$ cm$^{2}$s$^{-1}$ in the XY plane and about
$1.2\cdot\,10^{26}$ cm$^{2}$s$^{-1}$ in the Z direction for the experiments
IIIb, IIId and IVb (see Table 2). The computations checking the influence of
different physical magnetic diffusion values on the magnetic field
evolution were made in Paper I.
The experiments III and IV last $6\cdot\,10^{8}$~yr.

\begin{figure}
\resizebox{\hsize}{!}{\includegraphics[angle=-90]{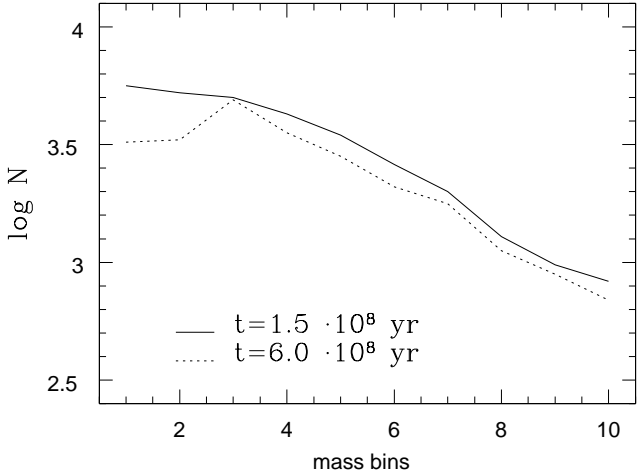}}
\caption[]{Number density of clouds per mass bin (logarithmic mass intervales)
  for the model III
at $t=1.5\times10^8$\,yr (solid line) and $t=6.0\times10^8$\,yr (dotted
line). Mass bin 1 is the lowest cloud mass class
(~5$\cdot\,10^2$ M$_\odot$); mass bin
10 is the most massive one (5$\cdot\,10^5$ M$_\odot$)}
\label{massbins}
\end{figure}

To compare the adopted numerical algorithms for the magnetic diffusion
term in the induction equation (\ref{eq2}) an initially random field is evolved
without any velocity field ($\vec{v}=0$).

\section{Pattern velocity and Lindblad resonances}

The non-axisymmetric structure in the disk has an angular pattern
speed given by the angular velocity $\Omega_p$. 
Three radii in each galaxy are of specific interest because
the pattern speed resonates with the eigenfrequencies of the
unperturbed galaxy. The first order epicyclic
theory (e.g. Binney and Tremaine 1987)
superimposes harmonic oscillations on the rotation in both
the radial and tangential directions with a characteristic
frequency called the epicyclic frequency
$\kappa= 2\Omega \left[1+{1\over 2}{d\, ln \Omega\over{d \,ln r}}
\right]^{1/2}$.
Thus, in the proper rotating frame, the particle will move in a
retrograde sense around a small ellipse, called an epicycle,
with axial ratio $\kappa/2\Omega$.
The resulting motion in the inertial frame is a rosette orbit, generally
not closed.
To the first approximation, the motion of a gas cloud or a star is thus
governed by two frequencies. In a frame corotating with a spiral these are
$\Omega(r)-\Omega_p(r)$ and $\kappa(r)$. When they are comparable, i.e.
when the relative frequency
$\nu=(\omega-m\Omega)/\kappa=m(\Omega-\Omega_p)/\kappa$
is equal to a rational number, we have a resonance.
The three main resonances in a galaxy (with a two-arm spiral pattern $m=2$)
are

- the inner Lindblad resonance (ILR) where $\nu=-1$ and $\Omega(r)-
\kappa(r)/2=\Omega_p$

- the corotation resonance (CR) where $\nu=0$ and $\Omega(r) =\Omega_p(r)$

- the outer Lindblad resonance (OLR) where $\nu=+1$ and
$\Omega(r)+\kappa(r)/2=\Omega_p$

At the Lindblad resonances, the disk and the
pattern (bar, spiral arms) intensively exchange energy, mass and angular
momentum.
The radial velocity, for example, changes its sign at CR. Inside CR, the
material flows inward and outside CR it moves outward. Due to the
energy and angular momentum transfer, the orbits of the particles change by
90 degrees at every resonance. As we shall see,
this simple picture is complicated by the fact
that a bar as well as a spiral system 
has its own resonances. Due
to the interaction of the particles with the non-axisymmetric structure,
the bar and spiral arms are slowed down, thereby moving the resonance radii
outward. Larger portions of the disk are then involved in
the radial inflow of the material. Since
the resulting disk velocity field, which we use as an input parameter for
our magnetic field calculation, reflects this dynamical behaviour,
we analyze the pattern velocity
and the radii of the Lindblad resonances and corotation in our simulated
galaxies.

\begin{figure}
     \resizebox{\hsize}{!}{\includegraphics*[angle=-90]{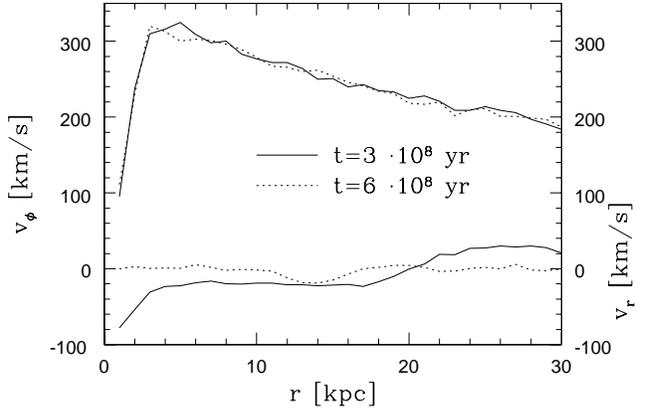}}
\caption[]{The rotation velocity and the radial velocity
at  $t=1.5\times10^8$\,yr (solid line) and $t=6.0\times10^8$\,yr (dashed
line) for the model III}
\label{vphivr}
\end{figure}
\begin{figure}
     \resizebox{\hsize}{!}{\includegraphics*[angle=-90]{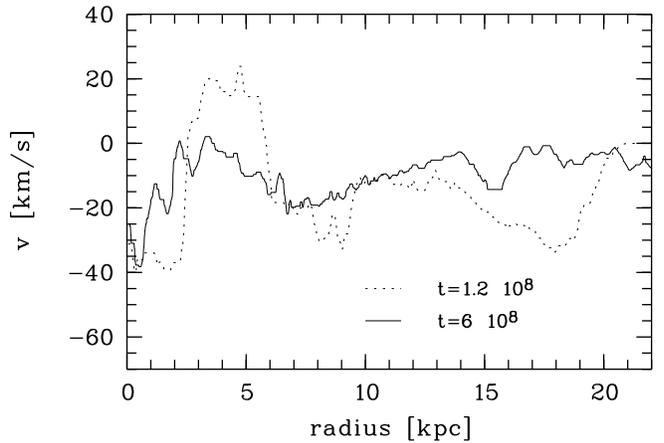}}
\caption[]{The radial velocity at a 1 kpc thick disk stripe ($z=0$)
at different time step are shown.}
\label{velo}
\end{figure}

Based on 
a Fourier analysis of the density distribution or of the gravitational
potential, which allows us to decompose
the perturbation present in the galaxy (Junqueira \& Combes 1996),
we are able to calculate the pattern velocity of our simulation at different
time steps. By plotting the pattern velocity $\Omega_p$ together with the
curves $\Omega \pm \kappa/2$ we determine the radii
of the resonances in the disk.

The uncertainty of the pattern velocity amounts to about $\pm$ 2\,km/s/kpc.
For the Lindblad radii the error increases from the inner to the outer
part of the disk.

\begin{figure*}
%     \resizebox{\hsize}{!}{\includegraphics[angle=0]{ml15.ps}}
\caption[]{{\bf a} Magnetic field vectors and the gas density
(grey plot) are plotted on the left for
the model IIIa at t= $1.5 \cdot\,10^8$\,yr.
The belonging intensity of the magnetic field (grey plot)
and the radii for the Lindblad resonances for the bar
(CR$_{\mbox{bar}}$ and OLR$_{\mbox{bar}}$, solid line)
and for the spiral system ( CR$_{\mbox{sp}}$, dashed line) are
shown in the right figure (the ILR$_{\mbox{sp}}$ are
at the same radius as the OLR$_{\mbox{bar}}$; the OLR$_{\mbox{sp}}$
are located outside the plot range). }
\label{mm15}
\end{figure*}

\addtocounter{figure}{-1}
\begin{figure*}
%     \resizebox{\hsize}{!}{\includegraphics*[angle=0]{ml25.ps}}
\caption[]{{\bf b} As in Fig. \ref{mm15}a at t = $2.5 \cdot\,10^8$\,yr. The
  radii of the Lindblad
resonances moved outwards during the simulation.}
\end{figure*}
\addtocounter{figure}{-1}
\begin{figure*}
%     \resizebox{\hsize}{!}{\includegraphics*[angle=0]{ml35.ps}}
\caption[]{{\bf c} As in Fig. \ref{mm15}a at t = $3.5 \cdot\,10^8$\,yr. In the
inner part of the disk a ILR$_{\mbox{bar}}$ appears.}
\end{figure*}
\addtocounter{figure}{-1}
\begin{figure*}
%     \resizebox{\hsize}{!}{\includegraphics[angle=0]{ml60.ps}}
\caption[]{{\bf d} As in Fig. \ref{mm15}a at t = $6.0 \cdot\,10^8$\,yr.
The resonances
for the spiral system are outside the plot range.}
\end{figure*}

%-------------------------------------------------------

\section{Results of Model III}

\subsection{3D N-body simulation}

Simulations of barred galaxies were performed by several
authors (see cf. Katz \& Gunn 1991; Gerin et al 1990;
Combes \& Elmegreen 1993; Norman et al 1996).
As in the present calculations, a strong spiral structure is formed
first. The two-armed spiral configuration becomes
more pronounced and more massive during the next ten time
steps. After $1.5 \cdot\,10^8$\,yr this structure
is clearly visible in the gaseous and stellar disks.

In comparison with model I (see Paper I),
experiment III exhibits a higher mass concentration
in the central part of the disk due to
the assumption of a bulge of smaller scale length.
This means that in this model, the bar appears earlier than in
case I (after 2.3 $\cdot\,10^8$\,yr).
A second reason for earlier bar formation is the
larger amount of gas in the disk (see also Friedli \& Benz 1993, Little
\& Carlberg 1991).
After $2.8 \cdot\,10^8$\,yr of evolution
the bar gets thicker, and during subsequnt time-steps the gas
concentrates along its major axis. A massive
elongated bar is formed which ends near the corotation radius, as in
the simulations of Sellwood (1981) and Combes \& Sanders (1981).

After $3.5\cdot\,10^8$\,yr the corotation radius of the bar is located
at 7.2\,kpc. The bar has a length of 14.2\,kpc and it concentrates 27\%
of the disk gas mass (further values are summarized in Tab\,\ref{tab3}).
In the outer part of the disk, a two-armed spiral pattern
is still visible (Fig.\,\ref{bari}). The pattern velocity
and the Lindblad radii are given in Table {\ref{tab3}}.

The orbits of the bar in the xy plane are similar to those of
the 2D-simulation of Sparke \& Sellwood (1987) and the 3D simulation
of Hasan et al. (1993). The high central gas concentration leads
to the appearance of $x_2$ orbits inside the ILR.
Due to the thickness of the bar, these orbits have a large population.
We show the isophotes of the stellar
disk density in Fig.\,\ref{iso}. It is clear that
these $x_2$ orbits create the isophote twist visible in the inner part
of the galaxy. The isophote twist
is also present in the simulations of Friedly \& Martinet
(1993) (see e.g. NGC\,1422) and it {\bf {cannot}} be interpreted as a
{\it{bar\,in\,bar}} (Shaw et al. 1993).

The mass spectrum of the molecular clouds changes during the evolution of
the disk, which is plotted in Fig.\,\ref{massbins} within 30 kpc of the disk.
After 6 $\cdot\,10^8$\,yr there are fewer particles in the
disk; in particular the number of low mass clouds decreases because of 
cloud-cloud collision.
The mass spectrum of the clouds in the bar is similar
to the mass spectrum of the total disk. However, the percentage
of extremely massive clouds in the bar is smaller than that in the disk.

During subsequent evolution, the pattern velocity of the
bar slowly decreases and the bar shrinks as in
other simulations, cf. Little \& Carlberg (1991).

\begin{table*}
\begin{flushleft}
\begin{tabular}{lllllllll}
\hline\noalign{\smallskip}
\hline\noalign{\smallskip}
timestep & $\Omega_{\mbox{bar}}$ & & ILR$_{\mbox{bar}}$
& CR$_{\mbox{bar}}$ & OLR$_{\mbox{bar}}$& & \\
& &$\Omega_{\mbox{sp}}$& & & = ILR$_{\mbox{sp}}$& CR$_{\mbox{sp}}$
& OLR$_{\mbox{sp}}$\\
\noalign{\smallskip}
[10$^8$\,yr] & [${\mbox{km}\over \mbox{s\,kpc}}$] & [${\mbox{km}\over \mbox{s\,kpc}}$]&[kpc]&[kpc]&[kpc]&[kpc]&[kpc]\\
\hline\noalign{\smallskip}
1.5 & 66.9 &21.9 & - & 4 & 6.5 & 13 & 18.9\\
2.5 & 57.3 &14.3 & - & 5 & 8.1 & 18.0 & 23\\
3.5 & 31.8 &12.2 & 2.1 & 7.2 & 11.5 & 24.1 & 27.3 \\
6.0 & 25.0 &11.0 & 3 & 9.9 & 17.9 & 29 & $\sim$ 35\\
\hline\noalign{\smallskip}
$\pm $& 2& 2 & 2.4 & 2.4 & 2.6& 3 & 4\\
\hline\noalign{\smallskip}
\hline\noalign{\smallskip}
\end{tabular}
\caption[]{\label{tab3}Pattern velocity of the bar and the
spiral system (sp) and the radius of
the corresponding resonances }
\end{flushleft}
\end{table*}

\subsection{The velocity field}
The velocity field of the disk given at different time steps is
the input for magnetic field calculation.
Fig. \ref{vphivr} presents the rotation curve and the average radial
velocity
at different time steps calculated in rings of constant distance.
Negative $v_r$ vectors are directed toward the galactic center.
Obviously,
there is mass transport toward the center, as 
might be expected.
Inside the bar, the radial velocity is dominated by the $x_2$-orbit
family  (Contopoulous \& Papayannopoulos 1980) which allows for
positive radial velocities. To examine the local radial velocity
in more detail, a cut in the disk at $z=0$ 
through the midpoint with a width of 1 kpc is made for two time steps
(Fig. \ref{velo}).
At 6 $\cdot\,10^8$\,yr the cut is taken along the bar.
This figure shows that the radial velocity reflects the
dynamical behaviour of the disk: e.g. at t=$6 \cdot 10^8 $yr
the ILR$_{bar}$ is located at 3 kpc and v$_r$ changes sign at
this radius.

\subsection{The 3D evolution of magnetic field: model III}

Four simulations of barred gas-rich galaxies have been performed,
charactaized by the inital configuration of the
magnetic field: a toroidal magnetic field
with (IIIb) and without magnetic diffusion (IIIa) and
a random field with (IIId) and without diffusion (IIIc)
(Table \ref{tab2}).
In Figs.\ref{mm15}a-d the evolution of the large-scale 
magnetic field for the case IIIb is presented. We show
only the case with non-zero turbulent magnetic diffusion
$\eta$. However, the computations
with $\eta=0$ result in similar structures (but with a higher
magnetic energy density, see Fig.\,\ref{ene}).
For the present calculations, the diffusion coefficient $\eta$
is about three times smaller than
in the case Id presented in Paper I, where diffusion significantly
influenced the magnetic field behaviour. The models in Paper I are also
affected by numerical diffusion. Thus, the present simulations
supersede those in Paper I. 

At the first evolutionary stage, the magnetic field
quickly responds to the velocity shear of the interstellar gas flow in
the non-axisymmetric gravitational disturbances.
Fig.\ref{mm15}a presents the magnetic field distribution at
$t=1.5 \cdot\,10^{8}$\,yr. The field vectors (left), shown in
the galactic plane, are superimposed onto the gas density grey plot (in
logarithmic scale). The magnetic field intensity distribution (right)
is plotted together with the circles of the Linblad resonances of the
the bar (CR$_{\mbox{bar}}$ and OLR$_{\mbox{bar}}$ solid line) and of the spiral arms (CR$_{\mbox{sp}}$ dashed line). From the beginning, the
magnetic field closely follows the dynamical spirals
resulting in a four-arm pattern visible in the figure. In one of the arms
the orientation of the magnetic vectors is opposite
to the initial
anti-clockwise direction of those in the toroidal magnetic field. This is a
result of the strong velocity shears present in this region
(see discussion in Sec.\,7). There are also two
regions with vector reversals located in the inner parts of the next two
arms,
close to the CR of the bar (see Fig.\ref{mm15}a).
The magnetic intensity distribution
(Fig.\ref{mm15}a, right) shows that the
magnetic features form nonuniform stripes
exhibiting generally spiral structure.
Fig.\ref{mm15}b presents the magnetic field configuration
at $2.5\cdot\,10^{8}$\,yr. The circles of the resonances for the bar
(CR $_{\mbox{bar}}$ and OLR$_{\mbox{bar}}$,
solid lines) and for the spiral arms (CR$_{\mbox{sp}}$, dashed line)
are superimposed onto the
intensity grey plot (right).
The dynamical structure shows the initial phase
of bar formation (see Fig.\ref{mm15}b) and two spiral arms.
The magnetic field is aligned
along the bar and spirals, but its distribution
is more complicated. In the central part of the right arm
(Fig.\ref{mm15}b, left),
the magnetic vectors are directed opposite to the vectors at the arm edges
(Fig.\ref{mm15}b, left), which is also visible in the intensity grey plot
(Fig.\ref{mm15}b, right). In the upper part of the figure, the region with
magnetic vector reversal is still present but the field is weaker
than in those magnetic features with an anti-clockwise direction of the vectors
(Fig.\ref{mm15}b, right). In Fig.\ref{mm15}c, we show the magnetic
field  distribution at $t=3.5\cdot\,10^{8}$\,yr. The ILR now
appears in the bar.
Three resonances (ILR$_{\mbox{bar}}$, CR$_{\mbox{bar}}$
and OLR$_{\mbox{bar}}$) for the bar are marked here by
solid lines. In the outer part of the disk,
the spiral arm system is still visible
(see Fig.\ref{bari}). The magnetic field vectors are again
aligned with these structures but also form additional arms and
regions with vector reversals. As in experiments I and II of Paper I,
the magnetic field vectors start to be mixed at both ends of the bar.
In Fig.\ref{mm15}c, magnetic spiral features form a pattern of four
non-uniform spirals (see Fig\ref{mm15}c, right).
However, there is now a considerable difference between the present and
$1.5\cdot\,10^{8}$~yr time step (Fig.\ref{mm15}a). In the previous evolutionary
stage, magnetic spiral features
closely follow the dynamical spirals. At $t=3.5\cdot\,10^{8}$\,yr, the
magnetic field has two more arms than the gas (Fig.\ref{mm15}c).
The regions with vector reversals are still present between the two left
magnetic spiral features, 
along the right side of the bar and in the inner part of the right
arm (Fig.\ref{mm15}c). A broad bridge of magnetic vectors connects the lower
end of the bar with the right side arm and it has no significant dynamical
counterpart.

{\bf The main conclusion from these facts is that
the magnetic field "remembers" the dynamical disturbances significantly longer than
they exist}. At $t=6\cdot\,10^{8}$~yr, the situation is even more evident.
The dynamical arms disappear but magnetic spirals remain
(Fig.\ref{mm15}d). The magnetic field lines are aligned with the bar
and create two diffused lanes consisting again of two sub-arms with reversed
magnetic vectors
(Fig.\ref{mm15}d, left). In the right arm,
the magnetic intensity is similar to stripes, both, with and without reversal
(Fig.\ref{mm15}d, right).

The evolution of the galactic magnetic field under the
influence of non-axisymmetric gas flows shows that the magnetic field
reacts
quickly even to weak disturbances in the field. The direction of
the magnetic vectors can easily be changed by velocity shear,
especially in the regions close to the Lindblad resonances  of the bar and
in the spiral arms. We could connect these field reversal
with the corotation radius
of the disk (see discussion in Sec.\,7). The
numerical simulations also show that magnetic field vectors are mixed
at the bar ends, as was shown in Paper I.

\begin{figure}
%     \resizebox{\hsize}{!}{\includegraphics{vrwsp15.ps}}
\caption[]{Topolines of the radial velocity superimposed
onto the magnetic
intensity grey plot for the simulations IIIb at $t=1.5\cdot\,10^{8}$\,yr}
\label{vrwsp}\end{figure}

In order to analyze the dependence of the
magnetic field intensity distribution on the velocity field shear,
Fig.\ref{vrwsp} presents the topolines of the radial velocity $v_{r}$ over the magnetic
intensity grey plot for the simulations IIIb at $t=1.5\cdot\,10^{8}$~yr.
The maximum of the magnetic intensity is located
in the region with the
strongest gradient of the $v_{r}$ component,
which coincides with the CR$_{\mbox{bar}}$ circle
(see Fig.\ref{mm15}a). On the inside of the main
magnetic spiral features (upper-central part of the figure)
the velocity distribution has a maximum, while,
on the outside, a minimum is visible
(see Fig.\ref{vrwsp}). Both maxima are distributed along the OLR$_{\mbox{bar}}$
resonance circle (Fig.\ref{mm15}a).
The next magnetic arm, which is seen in the lower part of the figure, is also
connected with such reversals (CR$_{\mbox{sp}}$ radius)
of radial velocity. The magnetic strength
of this arm is weaker than that of the previous one, but the
maxima and minima on both sides of the arm are not so extreme.
{\bf This supports the conclusion that the magnetic field reversals
appear due to the reversal of the velocity at the CR radii.}
The rest of the magnetic features are not so prominent and it is difficult to
associate them with changes in the velocity shear.

\begin{figure}
     \resizebox{\hsize}{!}{\includegraphics[angle=-90]{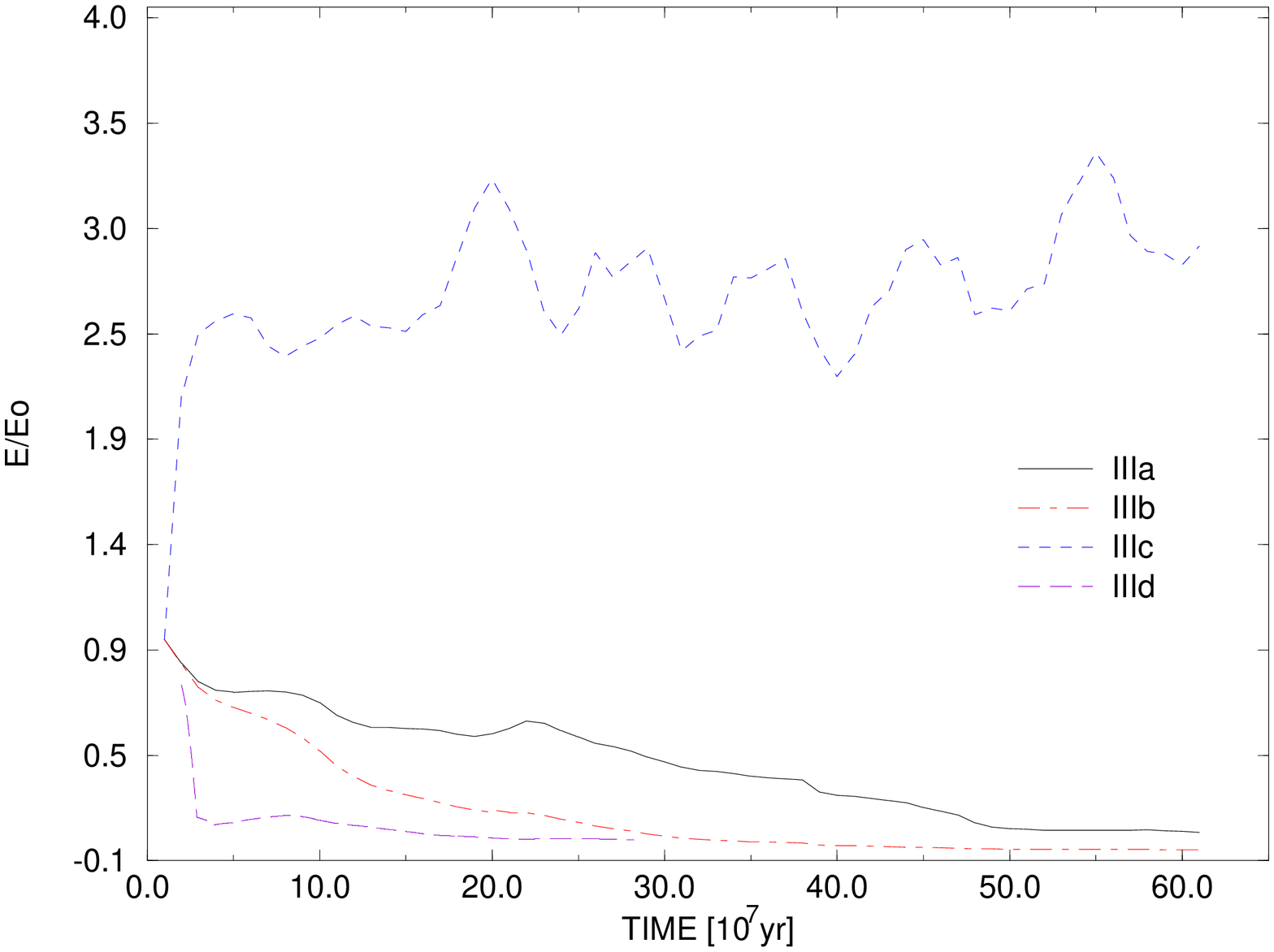}}
\caption[]{Time evolution of the magnetic energy density
$E$ normalised to the initial energy density $E_o$ for the models III.
The figure legend is in agreement with Table 2}
\label{ene}
\end{figure}

Fig.~\ref{ene} presents the time evolution of the magnetic energy
density (E) normalised
to the initial energy density (E$_o$) for the four cases in
model III. The parameters if the four vases are given in Table 2.
The energy density curves
show the time evolution of the toroidal (case IIIa and b) and
random (case IIIc and d) input magnetic field structure. Simulation IIIa,
with no physical diffusion (solid line),
shows a constantly decreasing magnetic energy density
due to numerical effects and the lack of strong velocity shears compared
to the barred galaxy case in Paper I. However, three small
maxima are still present (due to the spiral arms and
the bar formation). In model IIIb (dashed-dotted line)
the turbulent diffusion diminishes the magnetic energy density
much faster than in the previous case. The density for
simulation IIIc (short dashed line) with the random input field configuration
increases by a factor of 2.5 and then preserves the energy density value
during the whole evolution. This maybe explained by 
the random character of the initial magnetic field and/or the lack of
physical diffusion. The characteristic scale of magnetic field
variation has a length of one grid step,
so the numerical resistivity is not efficient enough to diminish the
field. 

The problem of the numerical resistivity was tested in several
experiments with different input random field and without
physical diffusion. The magnetic energy decreased when the input
random magnetic field possessed the scale of change larger than
one grid step. The random input magnetic field with magnetic vector
randomly distributed at each grid point avoid numerical resistivity.
This fact is probably connected with the numerical resistivity which
is not strictly a magnetic one (Hawley \& Stone, 1995). 
Stationary fields, for example, do not diffuse away.

All other experiments made with the random magnetic field and with the
turbulent diffusion exhibit a rapid decrease in the magnetic
energy density  which is present in case IIId (long dashed line).
The random magnetic vectors are distributed in all directions and, under the
influence of the diffusion, they easily interact, diminishing
the magnetic field strength.
The decrease of magnetic energy density in each experiment
strongly suggests that, in order to explain observed galactic fields,
a physical mechanism (e.g. a magnetohydrodynamical dynamo) is necessary
for magnetic field amplification on the
scale of the whole galaxy.

In order to compare the numerical schemes used in our experiments (the diffusion
is implemented directly into the code with the CT algorithm and with
the "pulsed-flow" method (see Paper I)), we perform
two experiments with random input magnetic field structure and with
no velocity field. Both methods give a similar slope of decreasing magnetic
energy density. However, the CT scheme
maintains a slightly higher magnetic energy density
than the "pulsed-flow" method. This is because
the "outflow" boundary conditions used in the calculations with the CT method,
which are otherwise the most stable for our simulations, exhibit
small magnetic monopole values at the boundaries.

%--------------------------------------------------------
\begin{figure*}
%     \resizebox{\hsize}{!}{\includegraphics[angle=0]{mg10.ps}}
\caption[]{{\bf a} Magnetic field vectors and the gas density
(grey plot) are plotted left for the model IVb at t= $1.0 \cdot\,10^8$\,yr.
The belonging intensity of the magnetic field (grey plot)
and the radii for the Lindblad resonances for the spiral system
(solid line) are shown in the right figure.}
\label{ml}
\end{figure*}
\addtocounter{figure}{-1}
\begin{figure*}
%     \resizebox{\hsize}{!}{\includegraphics[angle=0]{mt15.ps}}
\caption[]{{\bf b} As in Fig.\,\ref{ml}a now for t=$1.5\cdot
10^8$\,yr. The positions of Linblad resonances move outward.}
\end{figure*}

\section{Results of Model IV}

\subsection{3D N-body simulation for model IV}

Model IV also assumes a developing  two-armed spiral pattern.
This spiral structure transports angular momentum into the outer part of the
disk and, outside the corotation radius CR$_{\mbox{sp}}$,
the gas particles gain angular momentum.
Due to the assumption of a less massive disk in this model,
no bar appears in the disk
during the simulation.
The mass concentration in the inner part of the galaxy is maintained
during the evolution, but the arms become weaker and weaker
(after approximately $2.5\cdot\,10^8$\,yr).
A spiral structure is sharpest at the earliest stages of evolution
(see also the simulation in Elmegreen \& Tomasson 1993).
The vertical scale of the stellar disk changes during  evolution from
$\pm$ 0.5
kpc ($1\cdot\,10^8$\,yr)
to $\pm$ 1.5 kpc ($2.5\cdot\,10^8$\,yr), especially at the outer parts of the
disk. The pattern velocity and
the resonances of the spiral mode are calculated as described above.
At 1.5 $\cdot\,10^8$\,yr, we find the pattern velocity of the m=2
mode is about 25 km/s/kpc (see Table \ref{tab5}).
The spiral structure is visible for the next 4 $\cdot\,10^8$\,yr.

\begin{table}
\begin{flushleft}
\begin{tabular}{llllll}
\hline\noalign{\smallskip}
\hline\noalign{\smallskip}
timestep & $\Omega_{\mbox{sp}}$ & ILR$_{\mbox{sp}}$
& CR$_{\mbox{sp}}$ & OLR$_{\mbox{sp}}$ \\
\noalign{\smallskip}
[$10^8$\,yr] & [${\mbox{km}\over \mbox{s\,kpc}}$] &[kpc]&[kpc]&[kpc]\\
\hline\noalign{\smallskip}
1.0 & 28.6.9& 2.5 & 9.0 & 15\\
1.5 & 24.31 & 3 & 9.8 & 17.8\\
2.0 & 22.16 & 3.5 & 11.3 & 19.3\\
2.5 & 18.59 & 4.2 & 13.5 & 26\\
\hline\noalign{\smallskip}
$\pm$ & 2 &  4 & 4 & 4 \\
\hline\noalign{\smallskip}
\hline\noalign{\smallskip}
\end{tabular}
\caption[]{\label{tab5}Pattern velocity the
spiral system  and the radius of the resonances for model IV}
\end{flushleft}
\end{table}

\addtocounter{figure}{-1}
\begin{figure*}

%     \resizebox{\hsize}{!}{\includegraphics[angle=0]{mt20.ps}}
\caption[]{{\bf c} As in Fig.\,\ref{ml}a now for t=$2.0\cdot
10^8$\,yr.}
\end{figure*}
\addtocounter{figure}{-1}
\begin{figure*}
%     \resizebox{\hsize}{!}{\includegraphics[angle=0]{mt25.ps}}
\caption[]{{\bf d} As in Fig.\,\ref{ml}a now for t=$2.5\cdot
10^8$\,yr. The ring for the OLR is now outside the plot range.}
\end{figure*}

\subsection{The 3D evolution of magnetic field: model IV}

The evolution of the magnetic field in the spiral galaxy (model IV)
is caused by an efficient and fast
expulsion of the magnetic field from the central part of the galaxy and
is followed for $4.0\cdot\,10^8$\,yr.

The magnetic field vectors superimposed onto the mass density distribution
(grey plot, left)
and the magnetic field intensity distribution (grey plot, right)
shown together with the Lindblad radii of the m=2 spiral mode system
(solid circles) are presented in Fig.\ref{ml}a-c at four times:
1.0, 1.5, 2.0 and 2.5 $\cdot\,10^8$\,yr for the case IVb.

At $1\cdot\,10^8$\,yr (Fig.\ref{ml}a),
the magnetic field forms two diffuse spiral arms.
In the arms, the magnetic intensity is non-homogeneous. The areas
of stronger and weaker field density are visible,
though outside the arms the intensity distribution is more
uniform. The field is quickly expelled out of the central part of the
galaxy, which  possesses a strong mass concentration. This is the case
throughout the
whole evolution for $4.0\cdot\,10^8$\,yr.
At this early stage, the bifurcations of the magnetic spiral
features are observed.
The magnetic field vectors are aligned along the outer parts of the
visible arms and regions of magnetic field reversal become visible
(Fig.\ref{ml}a, left).
For $1.5\cdot\,10^8$\,yr, the magnetic spiral features
become more distinct and characteristic "islands"
of stronger fields are visible along them. Thus, the field is still
inhomogeneous
and is driven out from the galactic center. It forms a spiral pattern.
The magnetic field vectors are aligned along visible arms and reversals
are stronger and more clearly visible. Areas with
vectors in opposition to the initial toroidal field are again found
at the CR circle (see Fig.\ref{ml}a, right).
For $2.0\cdot\,10^8$\,yr, the field is driven out of the central
part of the galaxy,
and the magnetic spiral features are nonuniform.
They are clear and sharp at the beginning
and more diffused at their ends. They begin to merge
with each other so that the reversals almost disappear. The magnetic
spirals extend further out of the disk than their dynamical counterparts.
The magnetic field intensity starts to develop a ring at the corotation
radius. At $2.5\cdot\,10^8$\,yr, the spiral structure of the
magnetic arms disappears. The field forms a clear ring
outside the dynamical structure, still aligned with the CR circle 
(see discussion in Sec.\,7).
Its vectors are in the anticlockwise
direction and possess an almost azimuthal component (axisymmetric spiral
structure - ASS?). It is still inhomogenous with "islands" of
stronger field density. Very weak magnetic spiral features
appear in the center,
probably caused by the numerical artifacts.

\begin{figure}
     \resizebox{\hsize}{!}{\includegraphics[angle=-90]{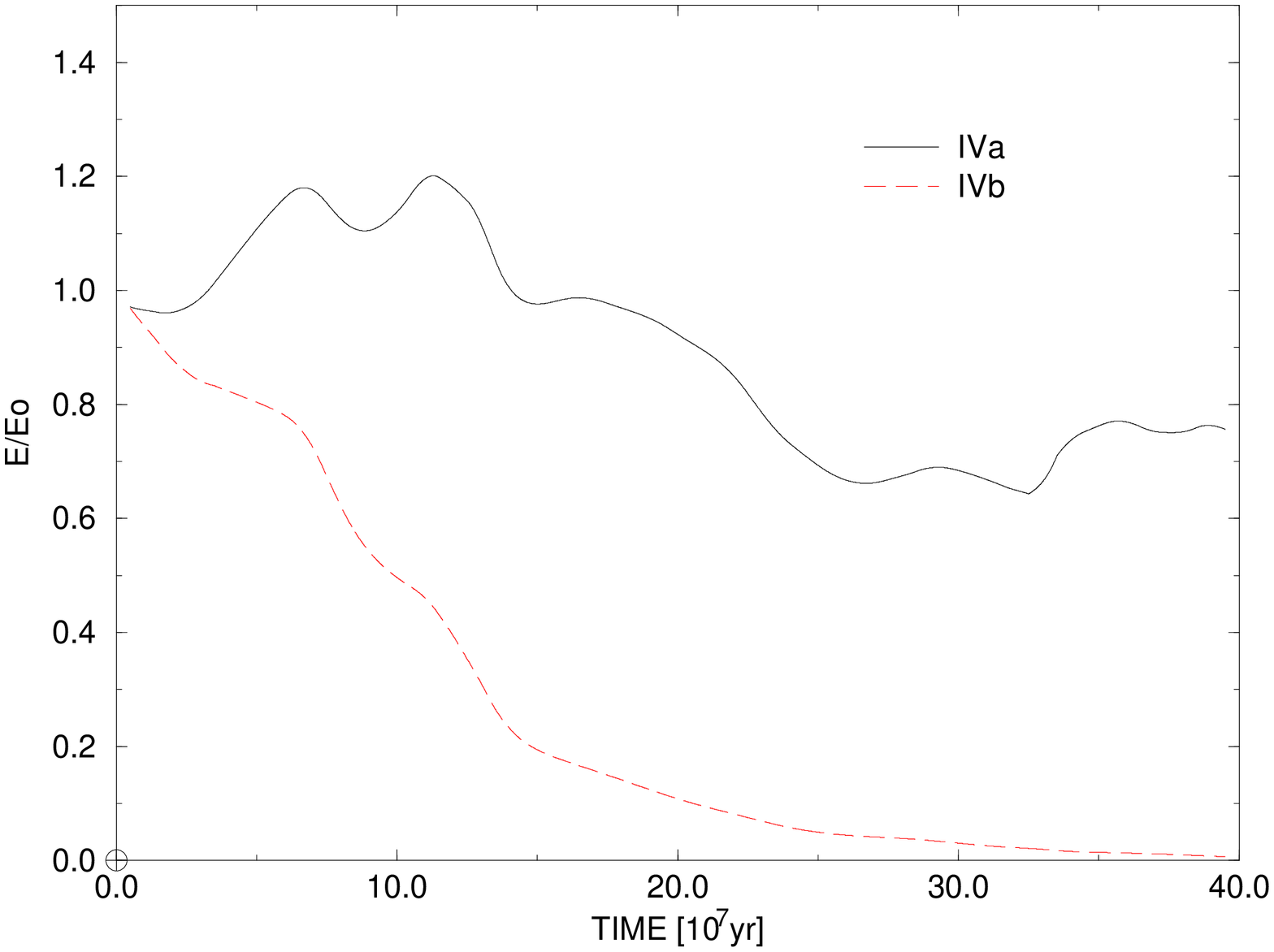}}
\caption[]{Time evolution of the magnetic energy density
$E$ normalised to the initial energy density $E_o$
for the model IV}\label{ene2}
\end{figure}

The magnetic energy density curves for model IV are presented
in Fig.\ref{ene2}.
The experiment with no physical diffusion (IVa, solid line)
results in two maxima of magnetic energy
density near $0.7 \cdot\,10^8$\,yr and $1.1\cdot\,10^8$\,yr. The first maximum
is connected with a strong velocity shear due to the initial
formation of spiral arms in the galactic disk. The second maximum is
the result of the velocity gradient caused by a strong mass transport
into the central part of the galaxy. The galaxy develops
a massive spiral structure that extends to the centre of the disk.
After two maxima, the energy density
decreases due to numerical resistivity, but still possesses further maxima
caused by continuous changes in the spiral pattern
(about $1.6 \cdot\,10^8$\,yr) and
later at about $2.9\cdot\,10^8$\,yr and $3.6 \cdot\,10^8$\,yr. These two
maxima are connected with the formation of a small bar in the central part
of the galaxy. The magnetic energy density for the case IVb with physical
diffusion (dashed line) decreases for the whole evolution showing
only small maxima at the same positions as in previous simulations.

The analysis of $B_{z}$, topoline maps made at different heights of the
galactic disk, shows that weak magnetic field waving is present in
both models (III and IV) (see Paper I for discussion).

\section{Conclusions and discussion}

A numerical model involving the three-dimensional and time-dependent
velocity of molecular gas is applied
to study the evolution of the large-scale  magnetic field under the
influence of non-axisymmetric dynamical structures.
We perform a comparison between bar and spiral arm activity to
investigate the effect of the different gas velocity fields.
The bar transports much more mass inward and angular momentum outward
than the disk with spiral arms alone.
Our present work demonstrates that the magnetic field is influenced more
strongly by the bar velocity shear than by spiral disturbances.
Our main conclusion is:
\begin{itemize}
\item[$\bullet$] The magnetic field dissipates much faster in spiral
than in barred galaxies due to the lower level of angular momentum transport.
\end{itemize}
The number of particles in the gas disk has no significant influence on 
magnetic field behaviour.

In all experiments, the regions with reversals of the magnetic vectors
appear due to velocity shears. We note that:
\begin{itemize}
\item[$\bullet$] Field reversals appear close to the corotation
radius
\end{itemize}
at which, according to standard theory (e.g. Athanassoula 1984),
the radial velocity of the gas flow changes its sign.
Outside the CR circle, the radial velocity
is directed outward while inside the CR, the gas flows inward if the
disk has only one resonance system. At the CR, there is a radial-velocity
minimum.
Since the magnetic field
is coupled to the gradients of the velocity field, it changes its direction
as well.
In addition to the corotation radius of the bars, there is also a
corotation radius of the spiral arms. Similarly, at the CR$_{\mbox{sp}}$,
the magnetic field changes its direction in response to
the varying velocity field at that radius. Such behaviour can be observed
in all our simulations but especially in model IV (spiral galaxy)
where the obtained ring of magnetic field (Fig.9c \& d) shifts together with
corotation radius of the spiral arms.

Changes in the velocity field at
the inner Lindblad resonance (ILR) are also visible in the magnetic
field.
The orbits change their sign by 90 degrees at every resonance.
Superimposed onto the global galactic rotation, this leads to a large-scale
spiral pattern, whose complicated velocity structure is traced by the magnetic
field via velocity gradients.

\,From our numerical experiments,
we demonstrate that due to a smaller number of Linblad resonances in spirals
\begin{itemize}
\item[$\bullet$] the regions of magnetic vector reversals are less
significant in spiral than in barred galaxies.
\end{itemize}

In addition, we have studied how the input magnetic
field affects the further development of the magnetic field distribution
and its intensity.
We have examined purely axisymmetric toroidal fields with an
anti-clockwise direction and random seed fields.

The case of non-axisymmetric input magnetic field has been
widely discussed in the paper Otmianowska \& Chiba (1995).
We analyzed the uniform protogalactic
magnetic field, distributed along the Y axis. For
all experyments, the magnetic field was quickly wound into the bisymmetric
structure, causing the initial growth of the magnetic energy.
As the time progresses the magnetic field in the inner part of the disk
was quickly swept out due to magnetic lines winding mechanism
and the magnetic energy always decreased very fast (after
$2\times 10^8 yr$).

In contrast to the computations with a toroidal magnetic field, those
with random  magnetic fields resulted in an
extremely rapid decrease in the magnetic energy density.
This suggests, that:
\begin{itemize}
\item[$\bullet$] Random seed magnetic fields in galaxies cannot explain
  the observed magnetic field in galactic disks.
\end{itemize}

As an interesting application, we note that
observations of the magnetic field in our Galaxy show
strong evidence for two reversals of the field vectors
(Vall$\acute{e}$e 1988, 1991, Han \& Qiao 1994). Vall$\acute{e}$e (1997) argues
that the region with magnetic vectors directed opposite to the
rest magnetic field in our Galaxy (which has a clockwise direction)
extends from
the Sagittarius arm (about $r=6.5$\,kpc) to the location of
the Sun (7.5 $\pm$ 1\,kpc).
{\bf These observed reversals of magnetic field are located in the
vicinity of the observed corotation radius in our Galaxy.}
Mishurov et al. (1997) estimated the pattern velocity of the spiral pattern
as $\Omega= 28.1 \pm 2.0$, and this sets the corotation radius of our
Milky Way at a radius of 7.2 $\pm $1.3 kpc.
This might confirm the statement that
the magnetic field vectors' distribution and direction depend
on the dynamical behaviour of the non-axisymmentric structures
and the positions of the Lindblad resonances as well.
Existing theoretical explanations (e.g. Poezd et al. 1993) concerning the
observed reversals in our Galaxy connect them with special conditions
of thin dynamo action and with structures inherited from seed
magnetic fields. Our results provide a simpler solution to this problem.
The regions of magnetic reversal appear due to
clear dynamical processes, closely connected with non-axisymmetric
features present in a galactic disk.
In a future work we will investigate whether there are more such
"coincidences" in other galaxies. For example, the ring structure
in model IV might be interesting for our neighbouring galaxy
M31 in which such a ring was detected 15 years ago by Beck (1982).

However the modeled ring is shifted further from the galactic center 
than the gaseous arms visible in Fig.9d. 
The observed magnetic ring coincides with the HI
torus at the radius of 10~kpc.

The field reversals are also interesting for comparison with the
observations of M81 by Krause et al. (1987a,b), who determined
a neutral line at which the magnetic field strength should vanish.

\begin{acknowledgements}
Part of this work was supported by the Deutsche Forschungsgemeinschaft
(Sonderforschungsbereich 328 {\it Evolution of Galaxies})
SvL thanks Prof. F. Combes for allowing us to use her code and for fruitful
discussions.
Our thanks to Detlef Elstner for his helpful comments and discussions.
KO wishes to express her gratitude to Marek Urbanik \& Marian Soida for
their valuable advice in the course of this project.
Calculations were
supported by the Forschungszentrum J\"ulich GmbH and
ACK CYFRONET-KRAK\'OW.
This work was partly supported
by a grant from the Polish Committee for Scientific Research (KBN),
grant no. PB/962/P03/97/12.
HL thanks Prof. R. Wielebinski and the
Deutsche Forschungsgemeinschaft for support (Grant LE 1039/2-1).
\end{acknowledgements}
%%%%%%%%%%%%%%%%%%%%%%%%%%%%%%%%%%%%%%%%%%%%%%%%%%%%%%%%%%%%%%%%%%%%%%%%%%%%
%%%%%%%%%%%%%%%%%%%%%%%%%%%%%%%%%%%%%%%%%%%%%%%%%%%%%%%%%%%%%%%%%%%%%%%%%%%%
%%%%%References
%%%%%%%%%%%%%%%%%%%%%%%%%%%%%%%%%%%%%%%%%%%%%%%%%%%%%%%%%%%%%%%%%%%%%%%%%%%%
%%%%%%%%%%%%%%%%%%%%%%%%%%%%%%%%%%%%%%%%%%%%%%%%%%%%%%%%%%%%%%%%%%%%%%%%%%%%

\end{document}